\documentclass[useAMS,usenatbib]{mn2e}
\usepackage{latexsym,graphicx,natbib,amssymb,array,aas_macros}

\newcommand\msun{\rm\,M_\odot}

\newcommand\rhpro{r_{hp}}

\newcommand\cmod{c_{\rm mod}}
\newcommand\tgas{\tau_M/t_{\rm cross}}
\newcommand\rhrt{r_{h,{\rm cl}}/r_{t,{\rm cl}}}
\newcommand\elow{\epsilon_{\rm low}}
\newcommand\rhfin{r_{h,\rm fin}}
\newcommand\rhecl{r_{h,\rm ecl}}
\newcommand\rhcl{r_{h,\rm cl}}
\newcommand\rhpd{r_{h,\rm PD}}

\newcommand\rhoecl{\rho_{\rm av,ecl}}
\newcommand\rhocl{\rho_{\rm av,cl}}

\newcommand\mfin{M_{\rm fin}}
\newcommand\mten{\mfin}
\newcommand\mecl{M_{\rm ecl}}
\newcommand\mcl{M_{\rm cl}}
\newcommand\mpd{M_{\rm PD}}
\newcommand\frh{f_r({\rm res. gas})}
\newcommand\fst{f_M({\rm res. gas})}

\newcommand\dgc{d_{\rm GC}}

\title[Initial conditions and assembly of old MW GCs]
      {Initial conditions for globular clusters and assembly of the old globular cluster population of the Milky Way}
      
\author[Michael Marks and Pavel Kroupa]
{
  Michael Marks$^{1,2,}$\thanks{Member of the International Max Planck Research School (IMPRS) for Astronomy and Astrophysics at the Universities of Bonn and Cologne; e-mail: mmarks@astro.uni-bonn.de (MM)} and Pavel Kroupa$^1$\\
  $^1$Argelander Institute for Astronomy, University of Bonn, Auf dem H\"ugel 71, 53121 Bonn, Germany\\
  $^2$Max-Planck-Institut f\"ur Radioastronomie, Auf dem H\"ugel 69, D-53121 Bonn, Germany\\
}       

\begin{document}

\date{Accepted ????. Received ?????; in original form ?????}

\pagerange{\pageref{firstpage}--\pageref{lastpage}} \pubyear{2009}

\maketitle

\label{firstpage}

\begin{abstract}
By comparing the outcome of $N$-body calculations that include primordial residual-gas expulsion with the observed properties of 20 Galactic globular clusters (GCs) for which the stellar mass function (MF) has been measured, we constrain the time-scale over which the gas of their embedded cluster counterparts must have been removed, the star formation efficiency the progenitor cloud must have had and the strength of the tidal-field the clusters must have formed in. The three parameters determine the expansion and mass-loss during residual-gas expulsion. After applying corrections for stellar and dynamical evolution we find birth cluster masses, sizes and densities for the GC sample and the same quantities for the progenitor gas clouds. The pre-cluster cloud core masses were between $10^5-10^7\msun$ and half-mass radii were typically below $1$ pc and reach down to $0.2$ pc. We show that the low-mass present day MF (PDMF) slope, initial half-mass radius and initial density of clusters correlates with cluster metallicity, unmasking metallicity as an important parameter driving cluster formation and the gas expulsion process. This work predicts that PD low-concentration clusters should have a higher binary fraction than PD high-concentration clusters.

Since the oldest GCs are early residuals from the formation of the Milky Way (MW) and the derived initial conditions probe the environment in which the clusters formed, we use the results as a new tool to study the formation of the inner GC system of the Galaxy. We achieve time-resolved insight into the evolution of the pre-MW gas cloud on short time-scales (a few hundred Myr) via cluster metallicities. The results are shown to be consistent with a contracting and self-gravitating cloud in which fluctuations in the pre-MW potential grow with time. An initially relatively smooth tidal-field evolved into a grainy potential within a dynamical time-scale of the collapsing cloud.
\end{abstract}


\section{Introduction}
\label{sec:intro}
Globular clusters (GCs) and old low-mass stars have often been used as local probes of Galaxy formation, since they preserve information about ancient times. In the past, the formation of the Milky Way (MW) has been investigated by means of kinematic studies of stars and their abundances \citep{CarolBeers07,bee02,bee01,els62}, as well as by horizontal branch morphology, metallicity and kinematic measurements of star clusters \citep{bekki07,mvdb05,mg04,z93,sz78}. In terms of the origin of the GCs, this led to a picture of Galactic formation in which GCs are divided in the old halo (OH) and young halo (YH) clusters. The OH clusters are located inside a Galactocentric distance of $\dgc\approx8-10$ kpc. Many of them appear to have formed coevally with the collapse of the protogalaxy \citep{els62,sw02,dea05,mf09}. The YH clusters have $\dgc\gtrsim8-10$ kpc and have been accreted over several Gyr \citep*{sz78}. All this information was gained from knowledge about the present day (PD) parameters of GCs. Knowing about the initial conditions at star cluster birth, however, would provide a deeper insight into the early formation processes of the MW since it would allow to probe directly the environment in which the GCs formed \citep*{g97,g98}.

Cluster masses at birth were larger than they are today since clusters suffer mass loss due to primordial residual-gas expulsion, stellar and dynamical evolution. And since the expulsion of gas leads to subsequent expansion, young and gas-embedded clusters were much more compact and denser than they are nowadays. Star clusters are the PD gravitationally bound remnants of these dense objects after the stars emerged from their natal cloud \citep*{tut78,kah01,bg06,bk07}. The overall change in the potential due to the gas loss leads to cluster expansion and the loss of stars, with the \textit{initial conditions} at the onset of this process deciding about cluster survival or destruction. The majority of the freshly hatched clusters are destroyed during this violent phase \citep*{ll03,bg06,gg08,b08,gb08} and their member stars become stars of the field. In this view, the OH GCs are the massive remnants of an initial population of embedded clusters that rapidly formed the population II halo of the MW \citep*{kah01,kb02,bkp08}. Some of them ended as bound clusters which, after stellar and two-body relaxation driven dynamical evolution, we can observe nowadays. So in order to understand physical properties of star clusters \textit{today} it is essential to understand how star clusters \textit{formed}, because the birth configuration determines the fate of a star cluster.

The time, $\tau_M$, over which the natal gas is removed from the cluster determines crucially whether a star cluster survives gas expulsion or not. \citet*{bkp08} provide an analytic formula to calculate $\tau_M$,
\begin{equation}
 \tau_M=7.1\times10^{-8}\frac{1-\epsilon}{\epsilon}\frac{\mcl}{\msun}\left(\frac{r_h}{\rm pc}\right)^{-1}\,\rm Myr,
 \label{eq:tau}
\end{equation}
based on the amount of energy needed to be put into the gas to overcome its potential energy. The deeper the potential, i.e. the larger the progenitor cloud mass, $\mcl$, the more difficult it is to remove the gas. A large star formation efficiency (SFE), $\epsilon$, leads to more and also more massive stars with stronger winds and radiation and there is less gas to remove. Finally, the larger the half-mass radius, $r_h$, the faster the gas is expelled since the overall potential is shallower for fixed $\mcl$ and $\epsilon$.

The gas expulsion time, thus, depends on the mass and the size of the cluster. From theoretical considerations mass and radius are related. Dependent on the exact form of the mass-radius relation of young star clusters, the gas expulsion time-scale, $\tau_M$, is affected more strongly or weakly. Virialised gas cores (radius $r_c$ and core mass $m_c$) are expected and observed to show a strong mass-radius relation, scaling as $r_c\propto m_c^{1/2}$ \citep*[i.e. constant surface density, e.g.][]{hp94}. Therefore the mass-radius relation of young clusters is expected to display a similar behaviour. If this relation is valid the gas removal time-scales would be essentially mass independent, $\tau_M\propto m_c^{1/12}$ \citep*{pf09}. In contrast, observations of young clusters do not show any significant mass-radius relation, $r_{cl}\propto m_{cl}^{0-0.1}$ \citep*[e.g.][]{l04,k05}, and the influence on the gas expulsion time-scale is a stronger function of mass, $\tau_M\propto m_c^{1/2}$ \citep*{pf09}.

Observationally determined SFEs lie between $\epsilon=0.2$ and $0.4$ \citep*{ll03}, so typically most of the mass remains in the gas to be expelled. $N$-body modelling showed that SFEs can be as low as $10-20$ per cent if the gas expulsion time-scales are sufficiently long and has to be at least $33$ per cent if the gas is lost instantaneously in order for a cluster to survive \citep*{lmd84,gb01,bk03a,bk03b,bk07}. SFEs may vary locally and could be different between high- and low-mass cores, which is also a viable explanation to wipe out the mass-radius relations observed in gas cores \citep*{az01}.

In Sec. \ref{sec:varimf} we present the observational data to be compared to the results of $N$-body experiments and in Sec. \ref{sec:models} we explain how we model the initial cluster size, mass and density. We derive the initial conditions and discuss correlations between and among initial and present day cluster parameters in Sec. \ref{sec:initcond}. In Sec. \ref{sec:galform} we connect the primordial conditions constrained in the former sections to develop a picture for the assembly of the old population of GCs located in the inner halo. Sec. \ref{sec:sum} summarises our main results.

\section{The De Marchi diagram and the metallicity-MF slope relation}
\label{sec:varimf}
\begin{figure}
 \begin{center}
  \includegraphics[width=8.3cm]{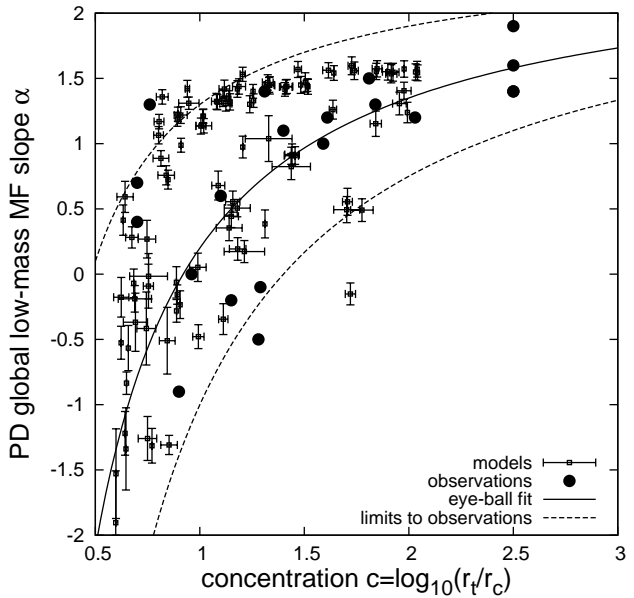}
 \end{center}
 \caption{Concentration parameter, $c=\log_{10}\left(r_t/r_c\right)$, vs. low-mass MF slope, $\alpha$, (the \textit{DMPP plane}). Weakly concentrated clusters are strongly depleted in low-mass stars and no cluster with a high concentration and a depleted MF is found (filled dots). This trend (black solid line, eq. \ref{eq:eyeball}) can't be understood in terms of purely secular dynamical evolution. However, $N$-body integrations of mass-segregated clusters at the time of the emergence from their birth molecular cloud with unresolved binaries in them (squares with error bars) reasonably reproduce the observed trend within the observational limits (dashed lines).}
 \label{fig:dmpp}
\end{figure}
The number of stars in the mass interval $m,m+dm$ is $dN=\xi(m)dm$, where $\xi(m)\propto m^{-\alpha}$ is the stellar mass-function (MF) with index $\alpha$. \citet*[hereafter DMPP]{dmpp07} showed that for Galactic GCs the value of $\alpha$ in the mass range $0.3-0.8\msun$ correlates with their concentration parameter $c=\log\left(r_t/r_c\right)$, i.e. the logarithmic ratio of tidal- and core-radius (Fig. \ref{fig:dmpp}, filled circles). The behaviour of $\alpha$ with $c$ cannot be understood to be the result of secular dynamical evolution without further assumptions \citep*[for details see DMPP;][]{mkb08a,bdmk08}. By fitting by eye DMPP proposed that the clusters obey a relation of the form
\begin{equation}
\alpha\left(c\right)=-\frac{2.3}{c}+2.5
\label{eq:eyeball}
\end{equation}
(solid black line in Fig. \ref{fig:dmpp}, hereafter referred to as the \textit{DMPP relation}).

\begin{figure}
 \begin{center}
  \includegraphics[width=8.3cm]{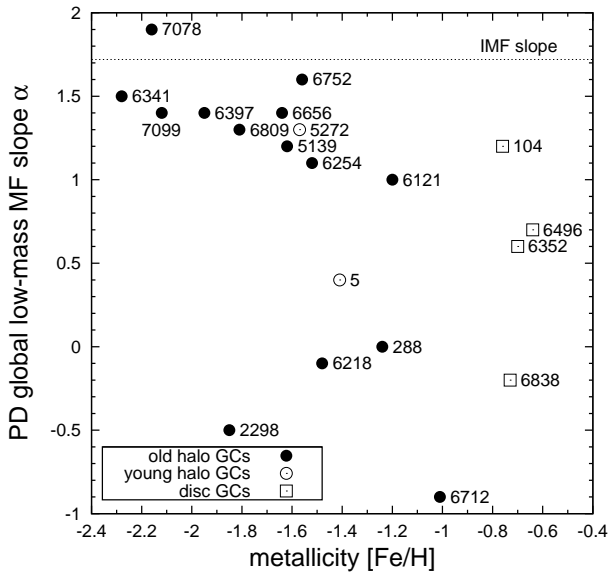}
 \end{center}
 \caption{Global cluster metallicity, [Fe/H] \citep*{h96}, vs. low-mass MF slope, $\alpha$, as in Fig. \ref{fig:dmpp}. Data points are labelled with the respective NGC or Pal catalogue number of the GC. The theoretical expectation would be that otherwise similar systems with a lower metallicity ought to have a bottom-light IMF compared to more metal-rich systems. Instead, clusters having a larger metallicity ($[{\rm Fe/H}]\gtrsim-1.5$) are depleted in low-mass stars ($\alpha\lesssim1$) giving support to the gas expulsion scenario and arguing against a variable IMF (see the text).}
 \label{fig:afeh}
\end{figure}
One exciting explanation for this trend is that it may signify for the first time evidence for a variability of the initial MF (IMF) at low masses. Low-$c$ clusters would form fewer low-mass stars than highly concentrated clusters. One would naturally expect this to depend on the metallicity, $[Fe/H]$, of the cluster. Fig. \ref{fig:afeh} displays the observed global MF slope versus the metallicity of the clusters in the sample. The anti-correlation between the two quantities is significant (see Tab. \ref{tab:spear} for a compilation of Spearman rank order correlation coefficients) and appears even slightly stronger when neglecting the four young, high-metallicity disc clusters NGC 104, 6352, 6496 and 6838. The data appear to suggest that stellar IMFs need to be flatter (smaller $\alpha$) in nowadays metal-rich clusters than in their metal-poor counterparts. In low metallicity environments, however, star formation should be terminated later by radiative feedback because the radiation couples less effectively to metal-poor gas. Stars would accrete matter over a longer time such that low-mass stars should be under-abundant \citep*{fa96}. Additionally, the Jeans mass is expected to be larger because of less effective cooling in gas clouds, again suggesting the IMF to have a larger average mass \citep*[e.g.][]{l98}. From these simple arguments we would expect to find depleted MFs in low-metallicity clusters, which is not observed. \textit{For these reasons the data in Fig. \ref{fig:afeh} are incompatible with our theoretical knowledge of star formation and up to now no purely star-formation-based explanation for this trend exists} \citep[see][]{mcc86,smcc87,dpc93}.

In an investigation of these issues, \citet*[hereafter MKB08]{mkb08a} showed, using $N$-body integrations of young, gas embedded clusters from \citet*[hereafter BK07]{bk07}, that the loss of the primordial residual-gas \textit{from initially mass-segregated clusters starting with a canonical IMF} \citep*{k01}\footnote{The canonical IMF has $\alpha=1.3$ for $0.08\leq m/\msun\leq0.5$ and $\alpha=2.3$ for $m/\msun>0.5$.} and \textit{containing unresolved binaries} (Fig. \ref{fig:dmpp}, squares with error bars) reasonably reproduces the trend within the observational limits (dashed lines).

\textit{Within the framework of gas expulsion we can now qualitatively explain the metallicity trend seen in Fig. \ref{fig:afeh}}: Stellar winds and the radiation of stars will lead to destruction of the gas cloud, in which the newly formed stars are embedded. Just as for the metal-dependend mass loss rates in stars \citep{mok07,vdkl01,kp00}, we suggest that the radiation of stars, which formed from a high-metallicity cloud, couples better to the gas through the dust, so that the residual-gas can be expelled more quickly via the efficient deposition of momentum into the interstellar medium. This leads to faster gas blow-out in high-metallicity clusters. In turn, the stronger change of the potential in a given time results in stronger expansion, i.e. a lower final concentration after gas expulsion, and stronger mass-loss over the tidal boundary, i.e. a lower value of $\alpha$. This is observed by DMPP.

However, cooling is more efficient in higher metallicity environments \citep{l98,fa96} and the emanating photon flux from the stars is higher if their atmospheres have a lower metallicity \citep*{kud02}. It is thus the interplay between the above three effects that determine the influence of metallicity on the gas expulsion process. We will demonstrate that metallicity affects the time-scale of gas expulsion (Fig. \ref{fig:taumetal} below) so that our idea in favour of a $[Fe/H]$-dependend gas loss process appears justified.

\textit{This supports the picture that residual-gas loss is indeed the driving mechanism to initiate the relation in Fig. \ref{fig:dmpp} and there may be no need to invoke a variable IMF at low masses}.

\section{The models}
\label{sec:models}
Our aim is to calculate the mass of the star forming cloud, $\mcl$, its size, $\rhcl$ and average density within the half-mass radius, $\rhocl$, each of the DMPP clusters must have formed from. This will enable us to investigate how GCs formed and what they and their birth sites looked like \textit{before} residual-gas expulsion.

For each GC we constrain a gas expulsion time-scale, $\tgas$, in units of the cluster initial crossing time, a star formation efficiency (SFE), $\epsilon$, and a value for the initial ratio of half-mass to tidal-radius, $\rhrt$, in Sec. \ref{sec:initcond}. These are the model parameters in BK07 which determine the fraction of final to initial half-mass radius, $\frh=\rhfin/\rhecl$, and the bound stellar mass-fraction, $\fst=\mfin/\mecl$, after the gas loss process (their figs. 1 and 4). Here, $\mecl=\epsilon\mcl$ is the mass in stars in the embedded cluster that formed out of the gas cloud and $\rhecl$ is the corresponding half-mass radius. We find a value for $\frh$ and $\fst$ for each of the 20 GCs in the DMPP sample by interpolating between the BK07 model grid.

We consider the PD $r_h$ and $M$ as the result of three independent evolutionary steps. The PD half-mass radius,
\begin{eqnarray}
 r_{h,\rm PD}&=&f_r(\begingroup \rm dyn.\;evo. \endgroup) \times f_r(\begingroup \rm st.\;mass\;loss \endgroup) \times \\
 &&\frh \times \rhecl\;,
 \label{eq:rhpd}
\end{eqnarray}
comes about from the change of the initial half-mass radius due to residual-gas expulsion, $\frh$, from mass loss due to stellar evolution, $f_r(\rm st.\;mass\;loss)$, and due to the dynamical evolution over a Hubble-time, $f_r(\rm dyn.\;evo.)$. We assume the latter to be negligible since the half-mass radius is approximately constant during dynamical evolution \citep*{kkb08}, so $f_r(\rm dyn.\;evo.)\approx1$ and eq. (\ref{eq:rhpd}) reduces to 
\begin{equation}
 r_{h,\rm PD}=f_r(\begingroup \rm st.\;mass\;loss \endgroup) \times \frh \times \rhecl\;.
 \label{eq:rhpdred}
\end{equation}
For these processes $f_r>1$ holds. According to \citet*{bm03}, clusters having a canonical IMF with stellar masses between 0.1 and 15 $\msun$ initially loose about $30$ per cent of their mass due to stellar evolution of the massive stars. In clusters with a fully populated IMF (up to 150$\msun$) this mass loss will be larger and closer to $40$ per cent. The loss of this gas from the cluster gives rise to further cluster expansion, which we assume to be adiabatic. In the case of slow expansion \citep*{h80} the PD half-mass radius is then given by
\begin{equation}
 \rhpd=\frac{\rhfin}{0.6}\;\Rightarrow\;f_r(\begingroup \rm st.\;mass\;loss \endgroup)\equiv\frac{\rhpd}{\rhfin}=\frac{1}{0.6}\;,
 \label{eq:adiabrh}
\end{equation}
where $\rhfin\equiv\frh\times\rhecl$ is the previously defined cluster half-mass radius after residual-gas expulsion and the factor $1/0.6$ accounts for the mass loss.

We do the same to constrain $\mecl$ and we include the effect of mass loss due to dynamical evolution. The PD stellar mass of a cluster is then
\begin{eqnarray}
 \mpd&=&f_M(\begingroup \rm dyn. evo. \endgroup)\times f_M(\begingroup \rm st.\;mass\;loss \endgroup) \times \\
 &&\fst \times \mecl\;,
 \label{eq:mpdred}
\end{eqnarray}
where $\fst$ is taken from the BK07 results as described above and $f_M<1$ for all processes. The product of the first two factors is
\begin{equation}
 \frac{\mpd}{\mten}=f(\begingroup \rm dyn. evo. \endgroup) \times f(\begingroup \rm st.\;mass\;loss \endgroup)\;,
\end{equation}
where $\mten=\fst \times \mecl$ is the above defined mass after residual-gas loss when the cluster has re-virialised. For $\mten$ we use the values derived in \citet{bdmk08}, who calculated initial cluster masses for the DMPP sample considering dynamical and stellar evolution only.

Via eqs. (\ref{eq:rhpdred}) and (\ref{eq:mpdred}) we can calculate the initial half-mass radius, $\rhecl$ and stellar mass, $\mecl$ for the young, gas embedded counterparts of the clusters in the DMPP sample. We then define the initial stellar density as
\begin{equation}
 \rhoecl\equiv\rhoecl\left(\leq \rhecl\right)=\frac{\mecl\left(\leq\rhecl\right)}{V\left(\rhecl\right)}\;,
 \label{eq:rhoav}
\end{equation}
i.e. the total stellar mass within $\rhecl$ divided by the volume of the sphere with radius $\rhecl$.

The corresponding quantities for the star forming cloud are obtained by replacing $\mecl$ with $\mcl$ and assuming $\rhcl=\rhecl$, i.e. stars and gas followed the same radial profile. From here on we set $\rhcl=\rhecl$. Note that the above initial densities refer to the point of time when the forming cluster is close to virial equilibrium before gas expulsion and approximately corresponds to the state of maximum density after cluster cloud-core contraction.

\section{Initial conditions}
\label{sec:initcond}
In order to derive the initial conditions described in Sec. \ref{sec:models} we first need to determine the time-scale of gas expulsion, $\tgas$, the SFE, $\epsilon$, and the strength of the tidal-field, $\rhrt$. In Figs. \ref{fig:rhrt} and \ref{fig:tgas} we compare the results of MKB08 with the observed data to constrain the initial $\rhrt$ and $\tgas$ values. MKB08 considered the effects of gas expulsion and unresolved binaries but they didn't include stellar and dynamical evolution in their analysis.

However, dynamical and stellar evolution do change $\alpha$ and $c$ and we will discuss their influence later (Sec. \ref{sec:discuss}). At this point we only note that the core-radius, $r_c$, entering the concentration parameter, $c$, is probably changed strongly by processes such as core collapse. The half-mass radius, $r_h$, is much more stable against dynamical evolution \citep{kkb08}. In Figs. \ref{fig:rhrt} to \ref{fig:eps} we therefore use a modified concentration parameter,
\begin{equation}
 c_{\rm mod}=\log_{10}\left(\frac{r_t}{\rhpro}\right)\;,
 \label{eq:modc}
\end{equation}
and use it instead of $c$ as a more robust measure. The radius $\rhpro\equiv\rhpd$ is the PD projected half-mass radius and $r_t$ is the PD tidal radius, both from \citet*[revision of 2003]{h96}. The tidal radius is assumed not to vary strongly after the cluster re-virialised after gas expulsion. The trend between $\alpha$ and $\cmod$ is similar to that in Fig. \ref{fig:dmpp} and enables us to gain more reasonable results.

\subsection{Tidal-field strength, time-scale of gas expulsion and star formation efficiency}
\label{sec:tauepsrhrt}
\begin{figure}
 \begin{center}
  \includegraphics[width=8.3cm]{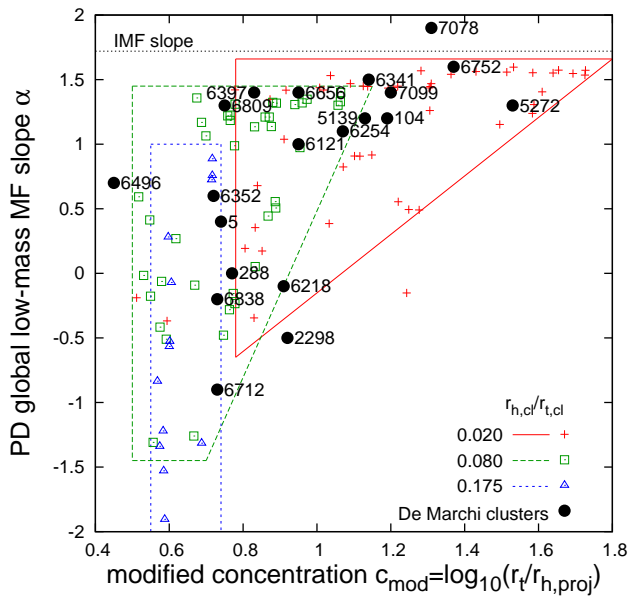}
 \end{center}
 \caption{As in Fig. \ref{fig:dmpp} but with the modified concentration parameter (eq. \ref{eq:modc}) on the abscissa. The theoretical data are coded for the initial tidal-field strength, $\rhrt$, in the simulations of BK07 with different symbols. Open symbols and crosses depict the $N$-body results, the filled circles labelled with their respective catalogue number are the data from the DMPP globular cluster sample. Quadrangles and triangles show areas with the same $\rhrt$. All initial models started from the dashed line, which is the slope of the canonical IMF \citep{k01} in the mass-range $0.3-0.8\msun$ ($\alpha\approx1.72$), in which the MF slope of the observed GCs was measured.}
 \label{fig:rhrt}
\end{figure}
From Fig. \ref{fig:rhrt} we find that the primordial tidal-field grows as $\cmod$ decreases. We grouped the initial tidal-field strength used in the models of BK07 into three groups. We draw boxes around areas containing (most of) the models with the same $\rhrt$-values by eye. Comparing the results of the integrations (open symbols and crosses) with the observational data (filled circles) we estimate the strength of the tidal-field for each DMPP cluster. If we can't assign a unique number to a GC, i.e. if a cluster is located in an overlap region, we choose a maximum and a minimum tidal-field strength defined by the overlapping regions and assign the mean value. NGC 6752, for instance, is assigned $\rhrt=0.02$, while for NGC 6656 $\rhrt=0.05$ was chosen, i.e. the average of $0.02$ (minimum tidal-field) and $0.08$ (maximum tidal-field). For GCs outside any of the regions (NGC 2298 and NGC 7078) the minimum and maximum tidal-field is chosen according to the $\rhrt$-values occurring around the same value of $\cmod$.

\begin{figure}
 \begin{center}
  \includegraphics[width=8.3cm]{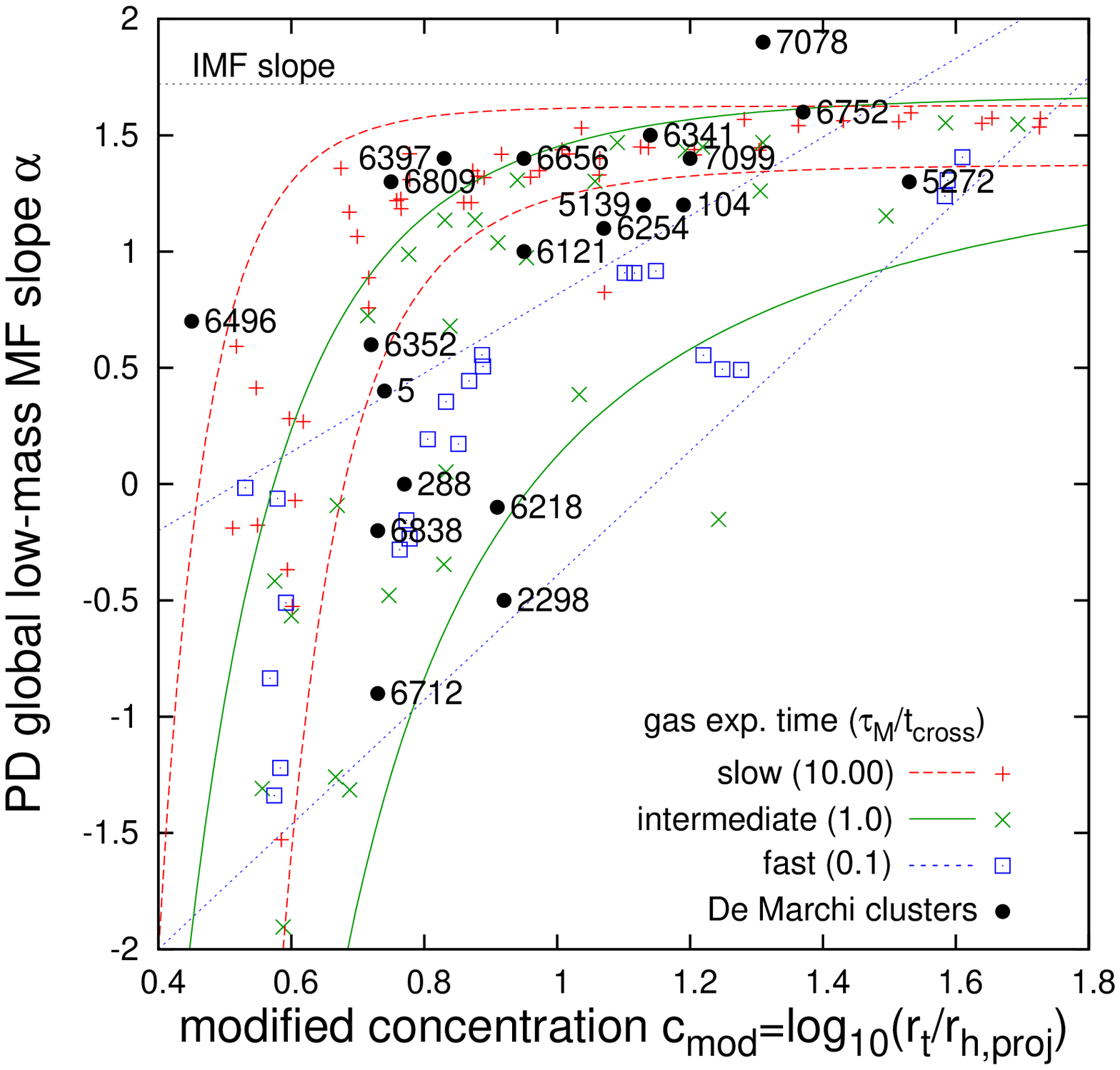}
 \end{center}
 \caption{Same figure and symbols as in Fig. \ref{fig:rhrt} but now coded for the residual-gas expulsion time-scale, $\tgas$. Two lines of the same type enclose regions with the same $\tgas$. Adiabatic ($\tgas\gg1$, dashed lines) and intermediate ($\tgas\approx1$,  solid lines) residual-gas expulsion times approximately follow the form of the DMPP relation, eq. \ref{eq:eyeball}. The trend for explosive gas removal ($\tgas\ll1$, short-dashed lines) is flat.}
 \label{fig:tgas}
\end{figure}
In Fig. \ref{fig:tgas} we divide the models into those with slow ($\tgas\gg1$), intermediate ($\tgas\approx1$) and fast ($\tgas\ll1$) gas removal. The location of a model cluster in the $\alpha-c$ plane is well described by a function of the form of the DMPP-relation, eq. \ref{eq:eyeball}, for slow and intermediate gas expulsion times. The trend is flat for fast gas removal. Areas in Fig. \ref{fig:tgas} containing models with similar gas expulsion times are enclosed by two lines of the same type. As for the tidal-fields we assign values for $\tgas$ to each cluster in the sample corresponding to their position in the DMPP diagram. E.g., NGC 6809 gets $\tgas=10$, NGC 5272 is assigned $\tgas=0.5$ ($\sim$the average of $0.1$ and $1t_{\rm cross}$) and, according to the models, NGC 6352 needed $5$ crossing-times to remove its residual-gas.

\begin{figure}
 \begin{center}
  \includegraphics[width=8.3cm]{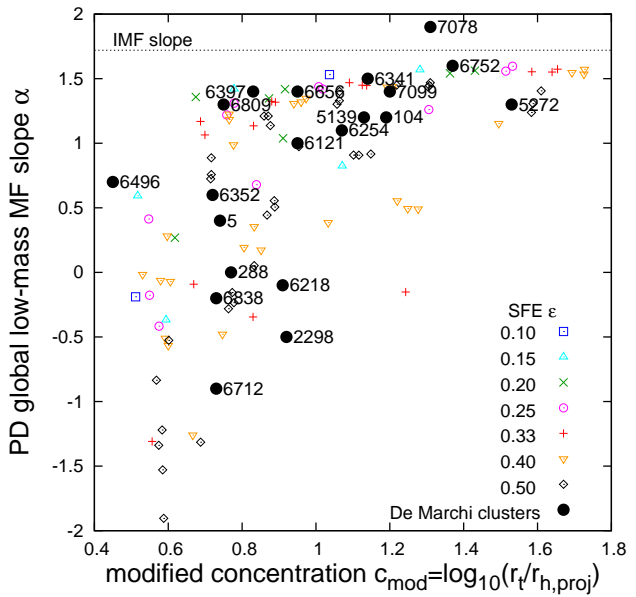}
 \end{center}
 \caption{Same picture and symbols as in Figs. \ref{fig:rhrt} and \ref{fig:tgas} but here coded for the SFE, $\epsilon$. SFEs of every value are distributed over the whole range of concentration except that for the depleted models ($\alpha\lesssim1$) mainly large SFEs ($\geq33$ per cent) are found. High SFEs for low-$\alpha$ clusters are needed to retain a bound cluster.}
 \label{fig:eps}
\end{figure}
The SFE shows no clear behaviour in the DMPP plane although it is an important parameter deciding about cluster survival or cluster destruction (Fig \ref{fig:eps}). For the models that are still closely resembling their IMF ($\alpha\gtrsim1$) basically all values from the SFE-range covered in the $N$-body integrations occur. Solely the strongly depleted ($\alpha\lesssim1$) model clusters nearly show large SFEs only ($\epsilon\gtrsim33$ per cent). For these reasons it is difficult to make a reasonable guess for the SFE for any of the DMPP clusters. We decided not to directly compare the observations with the models as before, but to look which SFE is at least necessary for a given parameter combination ($\rhrt,\tgas$) to retain a bound cluster. The SFE is then chosen such that at least four clusters from the grid of models of \citet*{bk07} around ($\rhrt,\tgas$) survive gas expulsion\footnote{Or two clusters, if either $\rhrt$ or $\tgas$ matches the model grid.}. Hence, we find lower limits to the SFE, $\epsilon_{\rm low}$.

\subsection{Size, mass and density}
\label{sec:smd}
For our modelling in Sec. \ref{sec:models} we can now directly get an estimate for $\rhcl$, $\mcl$ and $\rhocl$ with eqs. (\ref{eq:rhpdred}), (\ref{eq:mpdred}) and (\ref{eq:rhoav}) given the constraints on $\tau_M/t_{\rm cross}$, $\rhrt$ and $\epsilon$ found in the former section.

Note that since $\elow$ is a lower limit to the true SFE, and since the expansion factor, $\frh$, increases and the bound mass fraction, $\fst$, decreases with decreasing SFE (figs. 1 and 4 of BK07), the estimated value for $\frh$ is an upper limit, and $\fst$ is a lower limit compared to their true values. In turn, our estimates on $\rhcl$ and $\mcl$ have to be seen as lower and upper estimates, respectively.

\subsection{Results}
\label{sec:results}
\setlength\tabcolsep{4pt}
\begin{table*}
\begin{center}
\caption{Present day and constrained initial values for the clusters in the DMPP sample. Columns denote from left to right: Catalogue number and cluster type (disc, young/old halo), low-mass MF slope, $\alpha$, (modified) concentration, $c$ and $\cmod$ (eq. \ref{eq:modc}), tidal-field strength, $\rhrt$, the time-scale of gas expulsion, $\tgas$, the lower limit to the SFE, $\epsilon_{\rm low}$, the expansion factor and bound stellar mass fraction, $f_r\equiv\frh$ and $f_M\equiv\fst$, half-mass radii, $r_{h,x}$, masses $M_x$ and average densities, $\rho_x\equiv\rho_{{\rm av},x}$. The indices, $x$, denote present day (PD), initial stellar (ecl=stars only) and cloud (cl=stars+gas) values. $\mten$ is the mass after residual-gas expulsion from \citet{bdmk08}.}
\begin{tabular}{>{\footnotesize }l>{\footnotesize }c>{\footnotesize }c>{\footnotesize }c>{\footnotesize }c>{\footnotesize }c>{\footnotesize }c>{\footnotesize }c>{\footnotesize }c>{\footnotesize }c>{\footnotesize }c>{\footnotesize }c>{\footnotesize }c>{\footnotesize }c>{\footnotesize }c>{\footnotesize }c>{\footnotesize }c>{\footnotesize }c>{\footnotesize }c}
\hline
NGC & type & $\alpha$ & $c$ & $\cmod$ & $\rhcl/$ & $\tau_M$/ & $\elow$ & $f_r$ & $f_M$ & $\rhpd$ & $\rhcl$ & $\mpd$ & $\mten$ & $\mecl$ & $\mcl$ & $\rho_{\rm PD}$ & $\rho_{\rm ecl}$ & $\rho_{\rm cl}$ \\
 & & & & & $r_{t,\rm cl}$ & $t_{\rm cross}$ & & & & [pc] & [pc] & \multicolumn{2}{c}{$[10^5\msun]$} & \multicolumn{2}{c}{$[10^6\msun]$} & $[10^3]$ & \multicolumn{2}{c}{$[10^6\msun/pc^3]$} \\
\hline
104 & D & 1.20 & 2.03 & 1.19 & 0.02 & 1.0 & 0.25 & 4.46 & 0.47 & 3.65 & 0.49 & 7.00 & 11.0 & 2.35 & 9.40 & 1.72 & 2.36 & 9.45 \\
288 & OH & 0.00 & 0.96 & 0.77 & 0.08 & 0.5 & 0.40 & 1.98 & 0.39 & 5.36 & 1.63 & 0.48 & 2.20 & 0.57 & 1.43 & 0.04 & 0.02 & 0.04 \\
2298 & OH & -0.50 & 1.28 & 0.92 & 0.05 & 0.1 & 0.40 & 2.45 & 0.21 & 2.42 & 0.59 & 0.32 & 1.40 & 0.67 & 1.67 & 0.27 & 0.38 & 0.96 \\
Pal 5 & YH & 0.40 & 0.70 & 0.74 & 0.13 & 1.0 & 0.40 & 1.59 & 0.45 & 19.98 & 7.52 & 0.10 & 1.00 & 0.22 & 0.56 & $10^{-4}$ & $10^{-5}$ & $10^{-4}$ \\
5139 & OH & 1.20 & 1.61 & 1.13 & 0.02 & 1.0 & 0.25 & 4.46 & 0.47 & 6.44 & 0.87 & 15.0 & 27.0 & 5.77 & 23.08 & 0.67 & 1.06 & 4.22 \\
5272 & YH & 1.30 & 1.84 & 1.53 & 0.02 & 0.5 & 0.33 & 4.34 & 0.40 & 3.38 & 0.47 & 4.50 & 7.20 & 1.81 & 5.48 & 1.39 & 2.11 & 6.39 \\
6121 & OH & 1.00 & 1.59 & 0.95 & 0.05 & 1.0 & 0.25 & 3.32 & 0.32 & 2.33 & 0.42 & 0.67 & 7.00 & 2.19 & 8.78 & 0.63 & 3.50 & 14.01 \\
6218 & OH & -0.10 & 1.29 & 0.91 & 0.05 & 0.5 & 0.33 & 2.87 & 0.29 & 3.07 & 0.64 & 0.87 & 2.60 & 0.90 & 2.72 & 0.36 & 0.41 & 1.23 \\
6254 & OH & 1.10 & 1.40 & 1.07 & 0.05 & 1.0 & 0.25 & 3.32 & 0.32 & 2.31 & 0.42 & 1.00 & 2.70 & 0.85 & 3.39 & 0.97 & 1.39 & 5.54 \\
6341 & OH & 1.50 & 1.81 & 1.14 & 0.02 & 5.0 & 0.15 & 6.93 & 0.53 & 2.60 & 0.23 & 2.00 & 5.30 & 1.01 & 6.73 & 1.36 & 10.55 & 70.30 \\
6352 & D & 0.60 & 1.10 & 0.72 & 0.13 & 5.0 & 0.40 & 1.84 & 0.58 & 3.31 & 1.08 & 0.37 & 1.80 & 0.31 & 0.78 & 0.12 & 0.03 & 0.07 \\
6397 & OH & 1.40 & 2.50 & 0.83 & 0.05 & 10.0 & 0.15 & 5.38 & 0.61 & 1.55 & 0.17 & 0.45 & 1.60 & 0.26 & 1.76 & 1.44 & 6.11 & 40.72 \\
6496 & D & 0.70 & 0.70 & 0.45 & 0.08 & 10.0 & 0.25 & 3.18 & 0.66 & 6.25 & 1.18 & 0.82 & 2.20 & 0.33 & 1.33 & 0.04 & 0.02 & 0.10 \\
6656 & OH & 1.40 & 1.31 & 0.95 & 0.05 & 5.0 & 0.20 & 4.33 & 0.61 & 3.03 & 0.42 & 2.90 & 5.50 & 0.91 & 4.52 & 1.24 & 1.46 & 7.31 \\
6712 & OH & -0.90 & 0.90 & 0.73 & 0.13 & 0.5 & 0.50 & 1.40 & 0.47 & 2.75 & 1.18 & 0.94 & 5.10 & 1.09 & 2.17 & 0.54 & 0.08 & 0.16 \\
6752 & OH & 1.60 & 2.50 & 1.37 & 0.02 & 5.0 & 0.15 & 6.93 & 0.53 & 2.72 & 0.24 & 1.40 & 2.90 & 0.55 & 3.68 & 0.83 & 5.04 & 33.6 \\
6809 & OH & 1.30 & 0.76 & 0.75 & 0.08 & 10.0 & 0.25 & 3.18 & 0.66 & 4.53 & 0.85 & 1.10 & 3.30 & 0.50 & 1.99 & 0.14 & 0.10 & 0.38 \\
6838 & D & -0.20 & 1.15 & 0.73 & 0.13 & 0.5 & 0.50 & 1.40 & 0.56 & 1.87 & 0.47 & 0.17 & 0.84 & 0.15 & 0.30 & 0.31 & 0.17 & 0.34 \\
7078 & OH & 1.90 & 2.50 & 1.31 & 0.02 & 10.0 & 0.10 & 9.53 & 0.52 & 3.11 & 0.20 & 5.60 & 9.00 & 1.73 & 17.31 & 2.22 & 27.5 & 275.1 \\
7099 & OH & 1.40 & 2.50 & 1.20 & 0.02 & 5.0 & 0.15 & 6.93 & 0.53 & 2.67 & 0.23 & 1.00 & 2.60 & 0.50 & 3.30 & 0.63 & 4.78 & 31.85 \\
\hline
\end{tabular}
\label{tab:constraints}
\end{center}
\end{table*}
Given the DMPP relation and a universal low-mass IMF, the results of our constraints (Sec. \ref{sec:tauepsrhrt}) and the initial conditions for the DMPP GCs (Sec. \ref{sec:smd}) are as summarised in Tab.~\ref{tab:constraints}.

The estimated initial tidal-field strengths range from weak ($\rhrt\sim0.02$) to strong ($\rhrt\sim0.13$). The left-most part of Fig. \ref{fig:rhrt} ($c\lesssim0.7$) has only one observed cluster. This indicates that clusters forming in very strong tidal-fields may not survive for a Hubble-time and are quickly tidally disrupted \citep{bm03}.

\begin{figure}
 \begin{center}
  \includegraphics[width=8.3cm]{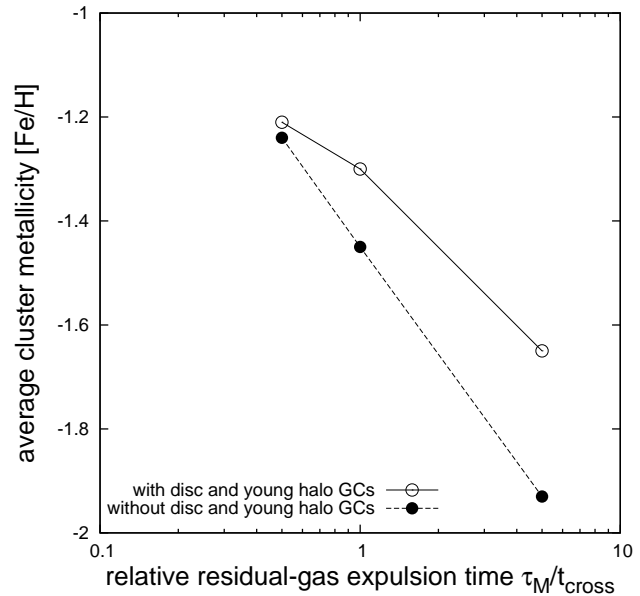}
 \end{center}
 \caption{The average metallicity for a group of clusters with similar constrained gas expulsion times is plotted against $\tgas$ (Sec. \ref{sec:results}). A clusters' metallicity determines whether the residual-gas is removed adiabatically (i.e. slow, $\tau_M>t_{\rm cross}$) or fast ($\tau_M<t_{\rm cross}$). Gas expulsion is more efficient in metal-rich systems, because the radiation driving the gas expulsion process couples more efficiently to metal-enriched gas, leading to larger expansion, mass-loss and, in turn, lower MF slopes (Sec. \ref{sec:varimf}, Fig. \ref{fig:afeh}).}
 \label{fig:taumetal}
\end{figure}
We find a large range of gas expulsion time-scales for the observed clusters, from essentially instantaneous ($\tgas\ll1$) to adiabatic ($\tgas\gg1$) gas removal, which have to be explained by the constrained initial conditions (mass, size \& metallicity). In Fig. \ref{fig:taumetal} we plot the average metallicity for groups of clusters with similar gas removal time-scales  ($\tgas>1,=1$ and$<1$, respectively; compare Tab.~\ref{tab:constraints}). We can thus confirm the idea from Sec. \ref{sec:varimf} that increasing metallicity reduces the time-scale of residual-gas expulsion and affects, in turn, the PDMF slope.

Estimates to the lower limits of the SFEs range from $\epsilon_{\rm low}\sim10$ to $\sim50$ per cent, so essentially all SFEs used in the $N$-body calculations occur. This covers observationally determined SFEs in young, gas embedded clusters which have SFEs of $20-40$ per cent \citep*{ll03}.

Most of the clusters share sub-pc initial half-mass radii (except Pal 5, NGC 288, NGC 6352, 6496 and 6712), suggesting that clusters form generally very compact, while Pal 5 stands out with a very large birth radius ($\rhcl=7.52$ pc, see discussion in Sec. \ref{sec:discuss}). Initially small clusters have also today small half-mass radii (Tab. \ref{tab:spear}). PD radii are $\sim3-10$ times larger than the birth values. Combining this with our constraints on the initial tidal-field strengths ($\rhrt$ in Tab. \ref{tab:constraints}), we find the initial tidal radii, $r_t$, to lie between $\sim5-60$ pc with an average of $\sim15$ pc only. Such small tidal-radii imply that clusters appear to have formed under very extreme conditions. A possible qualitative picture of Galaxy formation, in which such environments may have been frequent, will be discussed in Sec. \ref{sec:galform}.

Pre-cluster gas cloud-cores from which the stars of a cluster in our sample formed are estimated to have masses between $\sim3\times10^5\msun$ (NGC 6838) and $\sim2\times10^7\msun$ (NGC 5139, i.e. $\omega$Cen). Initial cloud masses thus span a range of two orders of magnitude. Present day massive clusters generally also had a massive progenitor cloud (Tab. \ref{tab:spear}). Cloud masses were on average about $30$ times larger than the PD cluster masses and about $\sim10$ times larger than the $\mten$-values. The stellar mass, $\mecl$, was $\sim10$ and $\sim2$ times larger than the PD and $\mten$ values, respectively. Pal 5, the cluster closest to dissolution (Sec. \ref{sec:discuss}), had a $\sim50\times$ more massive progenitor cloud. NGC 5272s progenitor on the other hand was only $\sim12\times$ more massive than today.

Initial average cloud densities range from $\sim10^2$ (Pal 5, has largest $\rhcl$) up to $\sim10^8\msun/$pc$^3$ (NGC 7078, has very low $\rhcl$ and may have survived a SFE of 10 per cent). Considering Pal 5 as an outlier, densities start at $\sim10^5\msun$ pc$^{-3}$. Thus, densities decrease from cloud formation to their PD stellar densities by factors lying typically between $\sim10^2$ (NGC 6712) to $\sim10^5$ (NGC 7078) with an average of $\sim2\times10^4$. The ratio of initial stellar to PD stellar densities is $10^2-10^4$ with an average at $3\times10^3$.

\subsection{Combination of initial cluster parameters}
\label{sec:combpara}
\begin{table*}
\begin{center}
\caption{Spearman rank-order correlation coefficients, $r_s$, between the quantities considered in the present paper (upper half). For no correlation, $r_s=0$, for perfect (anti-)correlation, $r_s=(-)+1$. The probability, $P(r_s)$, that the data show no correlation despite $r_s$ being different from zero, is given in the lower half of the table. We consider a relation to be significant, if $P(r_s)\leq0.02$. These numbers are bold-face in the lower part of this table. If numbers have a preceding '*', relations are considered to be trivial (e.g. mass and size are related to density, $\rho\propto M/r^3$) or not of interest (why should the PD density be connected with the initial half-mass radius?). The correlations having a bold-face number in the upper half are depicted with a figure in this paper. The correlations in italic style are discussed. The rows and columns denote: PD low-mass MF slope; $\alpha$; PD concentration, $c$; cluster metallicity, $[Fe/H]$; Galactocentric distance, $\dgc$ \citep[the latter two also from][2003 revision]{h96}; half-mass radii, $r_{h,x}$; logarithmic values ($\lg\equiv\log_{10}$) for the masses, $M_x$, and average densities $\rho_x\equiv\rho_{{\rm av},x}$ of the clusters from the DMPP sample. Indices as in Tab. \ref{tab:constraints}.}
\begin{tabular}{>{\footnotesize }l>{\footnotesize }r>{\footnotesize }r>{\footnotesize }r>{\footnotesize }r>{\footnotesize }r>{\footnotesize }r>{\footnotesize }r>{\footnotesize }r>{\footnotesize }r>{\footnotesize }r>{\footnotesize }r>{\footnotesize }r}
\hline
 & $\alpha$ & $c$ & $[Fe/H]$ & $\dgc$ & $\rhpd$ & $\rhcl$ & $\lg(\mpd)$ & $\lg(\mecl)$ & $\lg(\mcl)$ & $\lg(\rho_{\rm PD})$ & $\lg(\rho_{\rm ecl})$ & $\lg(\rho_{\rm cl})$ \\
\hline
$\alpha$ & 1.000 & \textbf{0.726} & \textbf{-0.647} & \textit{0.104} & \textit{-0.069} & \textit{-0.733} & \textit{0.646} & \textit{0.191} & \textbf{0.598} & \textit{0.109} & \textit{0.783} & \textbf{0.804} \\
$c$ &  & 1.000 & \textit{-0.556} & 0.273 & -0.405 & \textit{-0.848} & \textit{0.540} & 0.367 & \textit{0.680} & 0.450 & \textit{0.931} & \textit{0.911} \\
$[Fe/H]$ &  &  & 1.000 & -0.365 & 0.247 & \textbf{0.636} & -0.376 & -0.164 & -0.423 & -0.295 & \textit{-0.638} & \textbf{-0.657} \\
$\dgc$ &  &  &  & 1.000 & 0.077 & -0.133 & -0.002 & 0.111 & 0.125 & -0.024 & 0.192 & 0.167 \\
$\rhpd$ & &  &  &  & 1.000 & \textit{0.689} & 0.238 & 0.126 & -0.003 & -0.995 & -0.505 & -0.483 \\
$\rhcl$ &  &  &  &  &  & 1.000 & -0.335 & -0.131 & \textbf{-0.490} & -0.719 & -0.936 & -0.930 \\
$\lg(\mpd)$ &  &  &  &  &  &  & 1.000 & \textit{0.741} & \textit{0.870} & -0.187 & 0.545 & 0.566 \\
$\lg(\mecl)$ &  &  &  &  &  &  &  & 1.000 & 0.878 & -0.078 & 0.397 & 0.402 \\
$\lg(\mcl)$ &  &  &  &  &  &  &  &  & 1.000 & 0.059 & 0.719 & 0.729 \\
$\lg(\rho_{\rm PD})$ &  &  &  &  &  &  &  &  &  & 1.000 & 0.546 & 0.522 \\
$\lg(\rho_{\rm ecl})$ &  &  &  &  &  &  &  &  &  &  & 1.000 & 0.995 \\
$\lg(\rho_{\rm cl})$ &  &  &  &  &  &  &  &  &  &  &  & 1.000 \\
\hline
\hline
$\alpha$ & 0.000 & \textbf{0.000} & \textbf{0.002} & 0.663 & 0.771 & \textbf{0.000} & \textbf{0.002} & 0.421 & \textbf{0.005} & 0.649 & \textbf{0.000} & \textbf{0.000} \\
$c$ &  & 0.000 & \textbf{0.011} & 0.245 & 0.077 & \textbf{0.000} & \textbf{0.014} & 0.112 & \textbf{0.001} & 0.047 & \textbf{0.000} & \textbf{0.000} \\
$[Fe/H]$ &  &  & 0.000 & 0.113 & 0.295 & \textbf{0.003} & 0.102 & 0.490 & 0.063 & 0.207 & \textbf{0.002} & \textbf{0.002} \\
$\dgc$ &  &  &  & 0.000 & 0.748 & 0.577 & 0.992 & 0.640 & 0.600 & 0.920 & 0.416 & 0.482 \\
$\rhpd$ & &  &  &  & 0.000 & \textbf{0.001} & 0.313 & 0.596 & 0.990 & *\textbf{0.000} & 0.023 & 0.031 \\
$\rhcl$ &  &  &  &  &  & 0.000 & 0.149 & 0.581 & 0.028 & *\textbf{0.000} & *\textbf{0.000} & *\textbf{0.000} \\
$\lg(\mpd)$ &  &  &  &  &  &  & 0.000 & \textbf{0.000} & \textbf{0.000} & 0.431 & *\textbf{0.013} & *\textbf{0.009} \\
$\lg(\mecl)$ &  &  &  &  &  &  &  & 0.000 & *\textbf{0.000} & 0.743 & 0.083 & 0.079 \\
$\lg(\mcl)$ &  &  &  &  &  &  &  &  & 0.000 & 0.806 & *\textbf{0.000} & *\textbf{0.000} \\
$\lg(\rho_{\rm PD})$ &  &  &  &  &  &  &  &  &  & 0.000 & *\textbf{0.013} & *\textbf{0.018} \\
$\lg(\rho_{\rm ecl})$ &  &  &  &  &  &  &  &  &  &  & 0.000 & *\textbf{0.000} \\
$\lg(\rho_{\rm cl})$ &  &  &  &  &  &  &  &  &  &  &  & 0.000 \\
\hline
\end{tabular}
\label{tab:spear}
\end{center}
\end{table*}
In this section we look for possible relations between the constrained initial cluster parameters. In order to see if and where a correlation between the parameters exist we compiled Spearman rank-order correlation coefficients for the set of PD and initial cluster quantities used in this paper in Tab. \ref{tab:spear} (upper half). We present or discuss those correlations which have the largest significant Spearman coefficients (lower half of Tab. \ref{tab:spear}).

\begin{figure}
 \begin{center}
  \includegraphics[width=8.3cm]{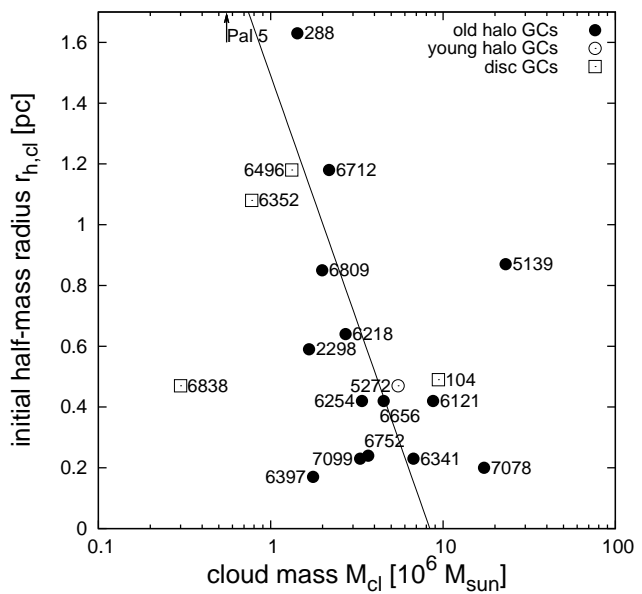}
 \end{center}
 \caption{Mass of the cluster forming cloud, $\mcl$, vs. initial half-mass radius, $\rhcl$. Different cluster types are indicated. The cluster Pal 5 has $\rhcl=7.52$ pc at the mass indicated by the upward arrow. Most GCs share sub-pc initial half-mass radii. The half-mass radius appears to decrease with increasing cloud mass. The solid line is to guide the eye.}
 \label{fig:rhmass}
\end{figure}
Young clusters do not show any mass-radius relation in observations \citep*{l04,k05} in agreement with a non-significant correlation between $\log(\mecl)$ and $\rhcl$. On the other hand their progenitor clouds are expected to show such a relation (Sec. \ref{sec:intro}), but we see only a weak (maybe not significant) correlation between initial cloud half-mass radius and cluster mass from Fig. \ref{fig:rhmass}. However, we note the improvement of the correlation when comparing $\rhcl$ with $\mcl$ instead of $\mecl$ which demonstrates the importance of comparing the relevant cluster parameters.

\begin{figure}
 \begin{center}
  \includegraphics[width=8.3cm]{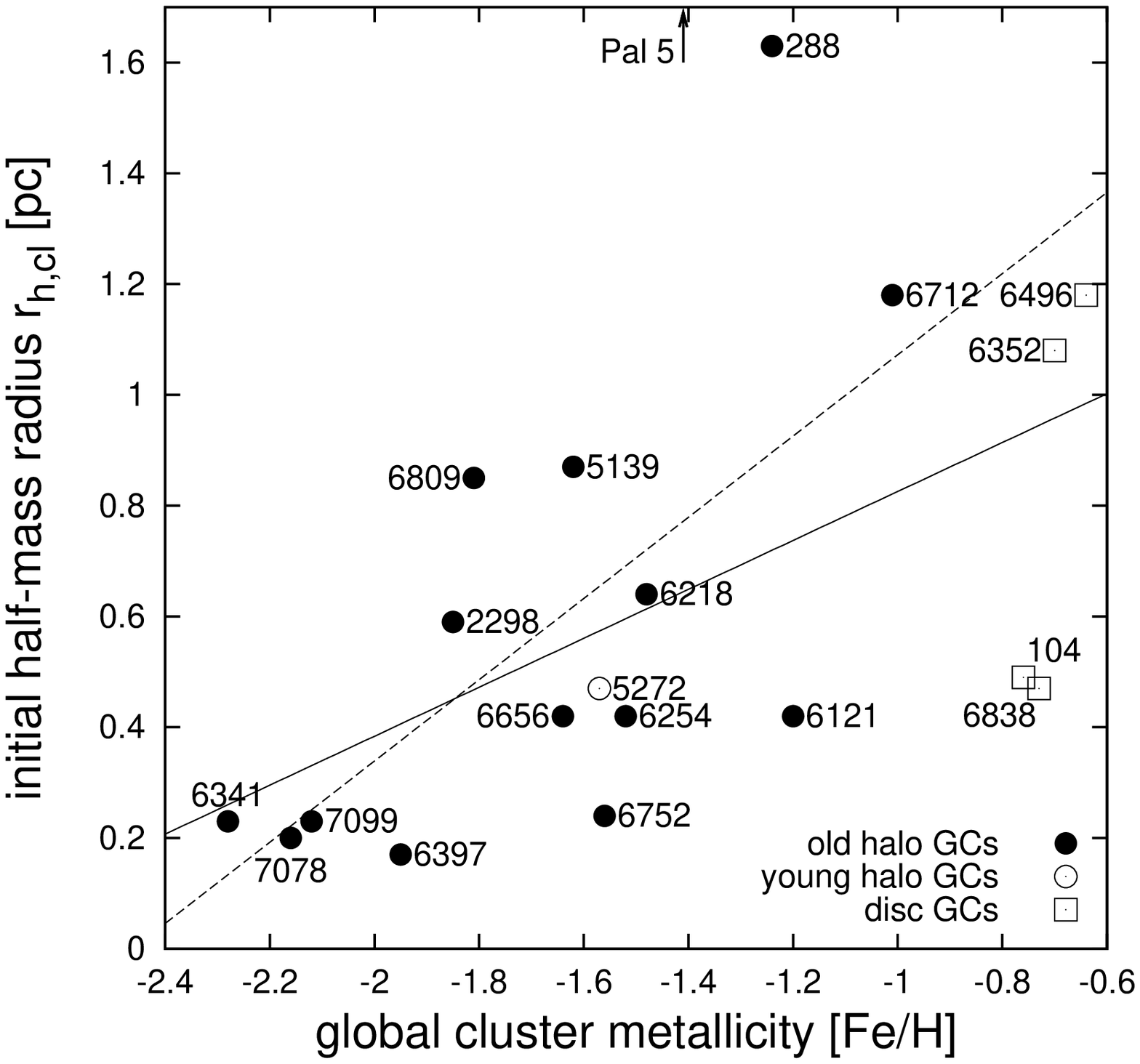}
 \end{center}
 \caption{The initial half-mass radius, $\rhcl$, increases with global cluster metallicity, [Fe/H]. The solid line presents a linear fit to all clusters in the sample. The dashed line is found excluding the disc and young halo clusters. The trend can be understood if metal-rich clusters cool more efficiently and fragment earlier into stars than metal-poor clusters, i.e. before global collapse of the star forming cloud occurs and also perhaps resulting from a feedback-regulated cluster formation process. Symbols as in Fig. \ref{fig:rhmass}.}
 \label{fig:rhmetal}
\end{figure}
We can understand the distribution of cluster size recognizing its dependence on metallicity (Fig. \ref{fig:rhmetal}). The initial half-mass radius appears to correlate with the cluster global metallicity in the sense that we find larger half-mass radii at larger metallicity. \textit{This trend can be expected, if higher-metallicity clusters can cool more efficiently and, thus, fragment earlier into individual stars. Metal-poor clusters have to wait for global collapse of the cluster to occur and a higher density before fragmentation becomes possible}. Also, if the cluster formation process is feedback-regulated, as suggested by the existence of the maximum stellar-mass $-$ star cluster-mass relation \citep*{wkb09}, then this correlation is naturally expected and would have a similar physical basis as metal-poor main-sequence stars being more compact than metal-rich stars.

\begin{figure}
 \begin{center}
  \includegraphics[width=8.3cm]{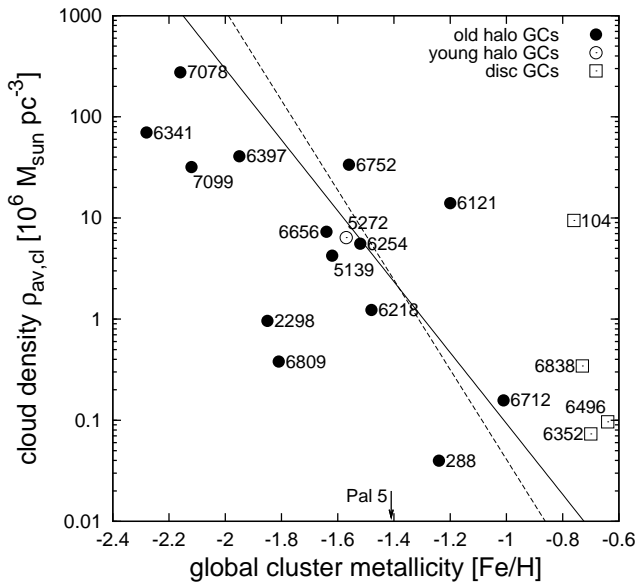}
 \end{center}
 \caption{Initial average cloud density within the half-mass radius, $\rhocl$, as a function of cluster metallicity, [Fe/H]. Pal 5 has $\rhocl\approx10^2\msun$ pc$^{-3}$ at the metallicity indicated by the downward arrow. Metal-poor clusters are denser than the relatively more metal-rich clusters as expected from our explanation for Fig. \ref{fig:rhmetal}. Symbols and best-fitting lines as in Fig. \ref{fig:rhmetal}.}
 \label{fig:densemetal}
\end{figure}
This view is supported when looking at initial cluster densities as a function of metallicity (Fig. \ref{fig:densemetal}). Given our explanation for Fig. \ref{fig:rhmetal} we would expect and indeed see that the initial stellar ($\rhoecl$) and cloud density ($\rhocl$) is larger in low-metallicity environments. The 2D hydrodynamical simulations of \citet*{hs09} confirm the easier fragmentation in more metal-rich environments and show that its dependence on metallicity is strong.

\begin{figure}
 \begin{center}
  \includegraphics[width=8.3cm]{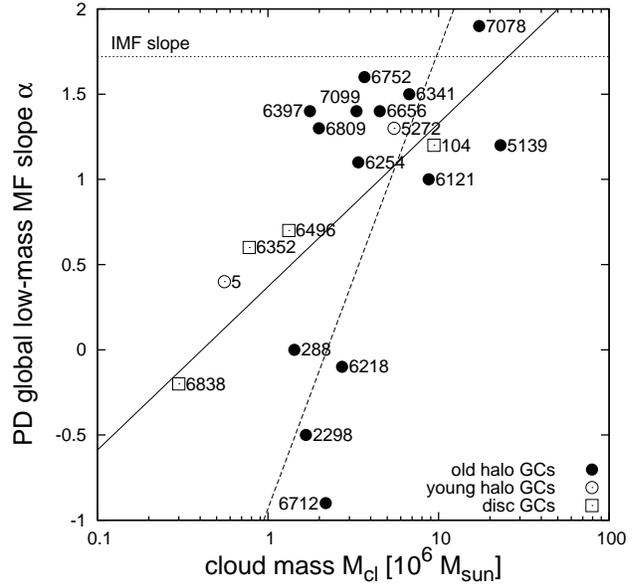}
 \end{center}
 \caption{The PD global low-mass MF slope, $\alpha$, is lower in star forming clouds with low masses, $\mcl$, as expected from eq. \ref{eq:tau} if residual-gas expulsion is the dominant source of low-mass star depletion. Symbols and best-fitting lines as in Figs. \ref{fig:rhmetal} and \ref{fig:densemetal}.}
 \label{fig:amass}
\end{figure}
Furthermore one expects gas expulsion to depend on the depth of the potential well the gas finds itself in (eq. \ref{eq:tau}), which is confirmed by Fig. \ref{fig:amass}. In a high-mass cloud the gas is removed less efficiently than from a low-mass one so the cluster retains more stars and a canonical $\alpha$. \textit{Again, we point out that we see a correlation only because we looked at the progenitor cloud mass.} The PDMF slope shows no relation with the total stellar mass forming out of the cloud material (Tab. \ref{tab:spear}).

The MF slope also correlates with metallicity. We have shown in Fig. \ref{fig:afeh} that the PDMF becomes more strongly depleted with increasing metallicity. Depleted and non-depleted observed clusters are distinctly apart in this diagram. It is expected that high-metallicity gas is more easily removed by radiation than metal metal-poor gas which ultimately leads to a PD depleted MF in initially mass-segregated clusters (Fig. \ref{fig:taumetal}).

\begin{figure}
 \begin{center}
  \includegraphics[width=8.3cm]{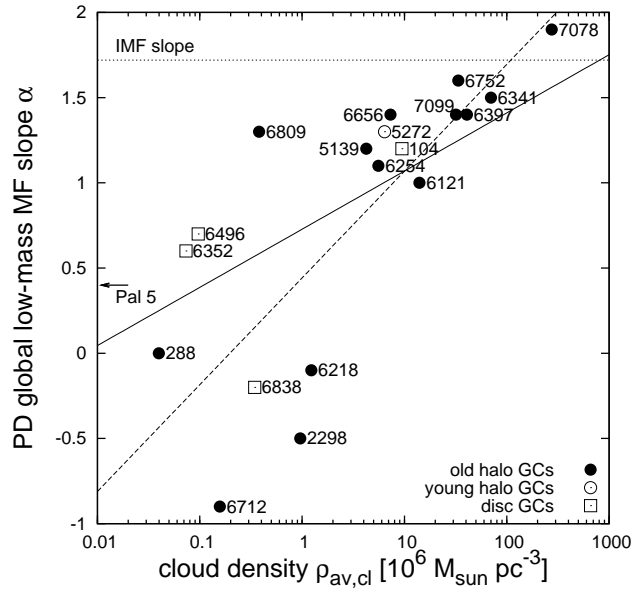}
 \end{center}
 \caption{Global low-mass MF slope, $\alpha$, versus the density of the star forming cloud, $\rhocl$. Pal 5 has $\rhocl\approx10^2\msun$ pc$^{-3}$ at the MF slope indicated by the leftward arrow. Low-mass star depleted clusters have the lowest concentration values today (Fig. \ref{fig:dmpp}) and are also initially the least dense objects. Symbols and best-fitting lines as in Figs. \ref{fig:rhmetal}, \ref{fig:densemetal} and \ref{fig:amass}.}
 \label{fig:adense}
\end{figure}
Low-mass star depleted clusters have the largest initial half-mass radii (Tab. \ref{tab:spear}) and, as a result, the lowest densities (Fig. \ref{fig:adense}). Large initial half-mass radii in low-$\alpha$ clusters will also lead to faster gas expulsion in them (eq. \ref{eq:tau}).

The PD concentration parameter, $c$, correlates with the same cluster parameters as $\alpha$ (Tab. \ref{tab:spear}), which is obvious given the DMPP relation (eq. \ref{eq:eyeball}). Thus, $c$ also correlates strongly with the initial densities, even more strongly than with the PD density\footnote{Because $c$ involves the radius ratio $r_t/r_c$ it may not trace density at all.}. This suggests that the PD $c$-value is a measure for the initial density.

\subsection{Discussion}
\label{sec:discuss}
The analysis above to constrain initial conditions of Galactic GCs yields remarkably nice results and information about how (globular) clusters might have formed.

Based on the still small sample of clusters in Figs. \ref{fig:afeh}, \ref{fig:rhmetal} and \ref{fig:densemetal}, \textit{metallicity appears to be an important parameter in determining the details of gas expulsion and cluster formation}. The existence of metallicity trends with initial parameters further suggests that the observed metallicities are primordial and clusters can't be strongly self-enriched. Metal-enrichment, however, occurs as seen by the multiple stellar populations observed in massive GCs. Next to NGC 5139 ($\omega$Cen, \citealt{VillPio09}) and NGC 6656 (M22, \citealt{lee09}), Terzan 5 in the Galactic bulge has recently been shown to host two stellar populations \citep{f09}. Metallicities for the different populations in $\omega$Cen and Terzan 5 differ by $\Delta[Fe/H]=0.5$ dex at most. They are both thought to be cores of former massive dwarf galaxies such that they were able to retain their supernova ejecta from which a second generation of stars formed \citep{lee09} and later merged with our Galaxy and can in that sense viewed to be exceptional. Alternatively, these clusters may have been immersed in a dwarf galaxy from which they repeatedly accreted interstellar matter which initiated the formation of new stars \citep*{pak09}. In M22 the metallicity difference is much less, $\Delta[Fe/H]=0.2$. Other clusters show virtually no dispersion in their iron content. That is, self-enrichment, if it did occur, did not significantly increase the bulk metallicity.

We found that the primary parameter determining $\alpha$ is the initial density (the Spearman coefficient is largest, Fig. \ref{fig:adense}). This results from a significant correlation of the PD $\alpha$ with the initial mass (Fig. \ref{fig:amass}) and a significant anti-correlation with the initial size (Tab. \ref{tab:spear}). The secondary parameter influencing the PDMF slope is metallicity (Fig. \ref{fig:afeh}). As explained, both dependencies point to a link between $\alpha$ and the details of residual-gas loss. A dependence of $\alpha$ on the PD density does not exist and might indicate that two-body relaxation driven evolution is not as important as residual-gas expulsion in determining $\alpha$.

A lower initial density of the presently low-$c$ clusters is also in concordance with the introduction of primordial unseen binaries in the analysis of MKB08: A binary fraction, $f_{\rm bin}$, was arbitrarily assigned according to the final $c$-value of the model clusters, i.e. after gas expulsion, with a larger value of $f_{\rm bin}$ at low concentrations in order to better match the models with observations. This is now naturally explained since the initially denser clusters will have dissolved more binaries because interactions happen more often \citep*{k95MN}. The prediction of this work would thus be that presently low-$c$ clusters ought to have a higher binary fraction.

Neither initial nor PD parameters correlate with the PD Galactocentric distance, $\dgc$, of the clusters (see Tab. \ref{tab:spear}), although a weak anti-correlation between $\dgc$ and $[Fe/H]$ may be suggestive (Sec. \ref{sec:galform}). While it is not clear why clusters should have formed at about the same distance where one sees them today, \citet*{dpc93} found that the low-mass MF slope, $\alpha$, is primarily determined by $\dgc$ (and the height above the Galactic plane), indicating that the strength and change of the tidal-field (tidal shocking) in determining $\alpha$ dominates over internal processes. We can not confirm their results since we don't find a dependence of the PD $\alpha$ on $\dgc$ with the DMPP sample of clusters. The second parameter determining the PD value of $\alpha$ in the analysis of \citet{dpc93} is metallicity, confirmed by our results. In their and subsequent works this dependence was not understood beside the idea that the correlation has its origin at cluster formation. We already qualitatively excluded a varying low-mass part of the IMF in dependence of $[Fe/H]$ to explain the DMPP relation since it is opposite to expectation (Sec. \ref{sec:varimf}) and we instead suggest a metallicity dependent residual-gas expulsion process to be the source of the depletion of low-mass stars in metal-rich, mass-segregated clusters (Figs. \ref{fig:afeh} and \ref{fig:taumetal}).

Pal 5 is outstanding in terms of its large initial radius and low initial density. This might be the case because it is closest to dissolution, i.e. the effect of significant dynamical evolution is important (see the next section). In this case the constraints might not be as good as for the other clusters. However, if the initial parameters for Pal 5 were as constrained in this work, this cluster might be the surviving remnant of a cluster complex \citep*[][see Sec. \ref{sec:galform}]{k98,fk02,bkf09}.

For each of the figures presented in the former section we do not find significant differences between disc, young and old halo clusters. The best-fitting lines change somewhat when excluding the disc and young halo clusters, but the improvement of the correlation measured by the Spearman coefficients isn't strong. Despite the likely different conditions experienced by halo and disk GCs, respectively, at the time of their formation, especially due to different tidal-fields, this holds as evidence that \textit{the mechanisms are similar in the different types of GCs}. As indicated by our constraints (Tab. \ref{tab:constraints}, Fig. \ref{fig:rhrt}), the disc GCs NGC 6352, 6496 and 6838 (except NGC 104) were born in stronger tidal-fields (larger $\rhrt$) than most of the halo GCs. By the response of gas expulsion to the primordial tidal-field of their more extreme host environments, these clusters find their PD position in the DMPP diagram.

Our constraints on the initial parameters rely, however, on the validity of the assumptions put into our analysis. The effects of dynamical and stellar evolution were included only in a simple way. A model taking gas expulsion \textit{and} these effects self-consistently into account would lead to improved results: The dispersion seen in the figures including the PDMF slope, $\alpha$, is relatively large (Figs. \ref{fig:afeh}, \ref{fig:amass} and \ref{fig:adense}) and could possibly be significantly reduced. In the following we will therefore discuss the relative importance of dynamical and stellar evolution.

\subsubsection{Dynamical evolution}
The quality of the constraints on the initial tidal-field strength, the gas expulsion time-scale and the SFE (Sec. \ref{sec:tauepsrhrt}) depends on how strong the fingerprint of residual-gas expulsion on the PD values $\alpha$ and $c$ is. Since MKB08 considered only the effect of residual-gas expulsion but didn't include stellar and dynamical evolution, the change of the parameters over a Hubble-time doesn't enter their models. Secular evolution driven by two-body relaxation alters the low-mass MF slope by the preferential loss of low-mass stars and the concentration parameter due to processes such as core-collapse. We accounted for the latter by using the modified concentration, $\cmod$ (eq. \ref{eq:modc}).

The ability of the $N$-body models of BK07 to reproduce the observations is a good starting-point to argue that residual-gas loss leaves a trace in the PD observables. This suggests that $\alpha$ and $c$ are established rather quickly after residual-gas loss. Indeed, we have shown that the PD concentration might as well be a direct result of the initial density (Tab. \ref{tab:spear}). And because of the dependence of PD $\alpha$ on $\mcl$ (Fig. \ref{fig:amass}), $\rhcl$ (Tab. \ref{tab:spear}) and $[Fe/H]$ (Fig. \ref{fig:afeh}), respectively, we are also confident that residual-gas expulsion is an important step in initiating the observed PD value of $\alpha$ although this is the shortest, but on the other hand most violent phase in the life of a cluster.

Further evidence is given by \citet{bdmk08} who explained the strongest low-mass star depleted clusters in the DMPP sample with secular evolution of mass-segregated clusters without taking residual-gas expulsion into account. They showed that the observed PDMF slope is additionally a function of the remaining life-time of a cluster and calculated that the clusters that are only slightly deficient in low-mass stars ($\alpha\geq1$) still need more than or about a Hubble time until complete cluster dissolution. These clusters can be considered as dynamically young objects and our constraints are probably best for them. The low-mass star depleted clusters ($\alpha\leq1$) have remaining lifetimes of $2-10$ Gyr with Pal 5 being closest to dissolution.

Following our results, the low-$\alpha$ clusters are the least dense objects initially (Fig. \ref{fig:adense}). After residual-gas loss (and stellar mass loss, see below) they will have strongly further reduced their density. Indeed, MKB08 already considered these objects as the fluffy remnants of residual-gas loss whose structural properties might not be altered strongly. So these clusters of the DMPP sample would possibly be least affected by dynamical evolution.

\subsubsection{Stellar evolution and mass-segregation}
As massive stars evolve they loose a fraction of their mass. When this mass is lost from the cluster, it will expand further changing its size and enhancing the loss of stars over the tidal boundary \citep[e.g.][]{bm03}. We included this effect by using the results of \citet{h80} for the change of the size (eq. \ref{eq:adiabrh}) and applying results from \citet{bdmk08} for the additional mass loss (eq. \ref{eq:mpdred}). \citet*{vmp09} showed that the effect of stellar evolution is even larger in mass-segregated clusters and that such clusters, if they fill their tidal-radius, would quickly dissolve under the influence of stellar mass loss. \citet{vmp09} conclude that clusters either can't have formed with a high degree of mass-segregation or they must be tidally-underfilled. Since gas expulsion in the models of BK07 naturally lead to tidally-filled clusters and MKB08 assumed strong mass-segregation, many of their model clusters would perhaps be quickly destroyed. Thus, the clusters observed today, as suggested by the results presented here, must have indeed formed highly compact and massive.

To summarise, a lot of work remains to be done to fully understand the evolution of star clusters from their birth to dissolution. Therefore self-consistent numerical modelling, including all dissolution mechanisms, i.e. residual-gas expulsion, stellar evolution and secular dynamical evolution, with realistic initial conditions (sizes and masses or equivalently, density, primordial binarity, ...) will be necessary, starting from the here derived initial conditions.

\section{Assembly of the Galactic old-GC halo}
\label{sec:galform}
In this section we combine the previously found results and develop a picture of the formation of the population II, i.e. the metal-poor and old halo of GCs.

We find that the gas expulsion time-scale, $\tgas$, and the initial strength of the tidal-field, $\rhrt$, vary more strongly between low concentration clusters and, in turn, increase the scatter in the MF slope, $\alpha$ (Figs. \ref{fig:rhrt} and \ref{fig:tgas}). Additionally, as the concentration, $c$, decreases, $\rhrt$ increases, i.e. PD low-$c$ clusters have formed in stronger tidal-fields. The tidal-field strengths found for the DMPP clusters can't be explained with the PD configuration of the MW globular cluster system, i.e. $\alpha$ and the Galactocentric distance, $\dgc$, show no correlation (see Tab. \ref{tab:spear}). We argue that all these observations can be understood to be the result of the assembly of the Galactic OH GC system within the initial cloud out of which the MW has formed.

The OH clusters appear to be co-eval to a good approximation \citep{sw02,dea05,mf09}. Despite their formation at the same time the clusters show a large spread in metallicity ($-2.4\lesssim[Fe/H]\lesssim-1.0$ for the OH clusters in the DMPP sample). In Sec. \ref{sec:discuss} we argued that the metallicity in Galactic GCs might be primordial and has not been significantly altered. If the metal content of the more metal-rich old population clusters comes from the products of stellar evolution which have been produced by earlier forming massive stars in other clusters, the inter-cluster medium (ICM) will be successively enriched with time. This material will then be recycled in later forming clusters. In this case a higher metallicity as observed for the low-$\alpha$ GCs (Fig. \ref{fig:afeh}) suggests that these clusters formed from an ICM enriched in metals and they should thus be somewhat younger than the relatively more metal-poor clusters. The age difference can then be at most a few hundred Myr only, corresponding to a dynamical time-scale.

\begin{figure}
 \begin{center}
  \includegraphics[width=8.3cm]{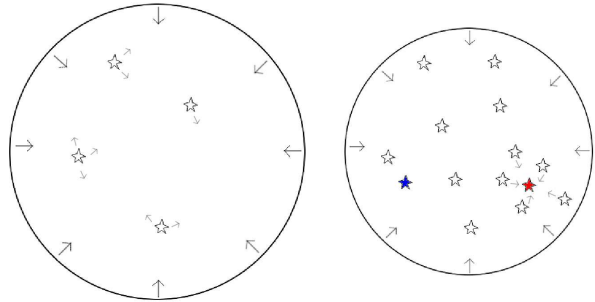}
 \end{center}
 \caption{Contraction scenario as described in Sec. \ref{sec:galform}. \textit{Left-hand picture:} In a collapsing cloud the first clusters (the stars in the picture) form all over the cloud in a smooth potential. The conditions are similar for all of them leading to comparable MF slopes and concentrations after gas expulsion. These clusters enrich their immediate environment with metals. \textit{Right-hand picture:} As the collapse proceeds more clusters form and the potential becomes grainy enhancing the mean tidal-field strength. Star clusters, e.g. the right filled cluster, forming near other clusters or dense clouds experience extreme conditions. They are more strongly enriched in metals from the surrounding objects than other clusters, e.g. the left filled cluster which is located in a more isolated place of the grainy potential, and they experience stronger tidal-fields than the first-forming clusters. This leads to strong differences between the formation sites and, in turn, to variations in the PDMF slope.}
 \label{fig:galform}
\end{figure}
These ideas lead us to a picture of the formation of the old inner and co-eval Galactic GC system, which might have formed during the contraction of a giant gas cloud out of which finally, after an additional long period of ongoing accretion and merging of formerly extragalactic systems \citep{sz78}, the MW has emerged \citep[possibly along the lines originally proposed by][]{els62}. In situ formation of the old GCs is in concordance with the age determination by \citet{mf09} who find that ``the age dispersion of the old group of globular clusters is not in contradiction with the formation from the collapse of a single protosystem''. \citet{zol09} find the radial transition between an in-situ and accretion-dominated halo in MW like galaxies to lie around $\dgc\approx20$ kpc, but most of the DMPP cluster even have $\dgc<10$ kpc (except NGC 288, 2298, 5272, 7078 and Pal 5). \citet[and citations therein]{CarolBeers07} argue that the dissipational merging of massive sub-galactic clumps builds up the inner halo which is followed by a stage of adiabatic flattening owing to the growth of the disk in order to explain the high-eccentricity orbits and higher peak metallicities observed for stars of the inner halo. While post-merger star formation may continue in the latter phase, also GCs will form in situ within the initial galaxy. \citet*{forbes10} point out that at least 1/4 of the Galactic GC system are of extragalactic origin and that, although OH clusters are commonly thought to have formed in situ, some clusters with OH classifications might actually be accreted.

However, whether still most the old halo GCs formed in situ, or whether the inner GC system is the result of the merging of sub-galactic building blocks \citep*[e.g.][]{bc01}, doesn't affect our discussion below. If the presented ideas are correct, we expect the scenario described there to be valid whereever the Galactic GCs formed, either during a single collapse or a collapse within the potentials of substructures which later merged, since the physical processes are fundamental. This can be supported by the two young halo clusters NGC 5272 and Pal 5, which possibly became part of the Galaxy during an accretion phase and fit perfectly well into the DMPP diagram (Figs. \ref{fig:rhrt}-\ref{fig:eps}) suggesting formation processes similar to those of old halo clusters and depending on the environment only (Sec. \ref{sec:discuss}).

In this frame, clusters which are depleted of low-mass stars, i.e. those that reside in stronger tidal-fields (Fig. \ref{fig:rhrt}) and expel their gas faster (Figs. \ref{fig:tgas} and \ref{fig:taumetal}) because they are relatively more metal-rich (Fig. \ref{fig:afeh}), were born at a later stage of (the GC's host-)galaxy formation. The environment in which the clusters formed must then have changed drastically within a short time.

The initial conditions can be understood, if the overall potential was rather smooth in the beginning. The first GCs formed all over the cloud experiencing similar smooth tidal-fields more or less independent of $\dgc$ and explaining the comparable values of $\rhrt$. The global metal content was still low, so one finds generally moderate to slow gas expulsion times (with respect to the crossing-time, Fig. \ref{fig:taumetal}) and the clusters were smaller, i.e. more compact initially (Fig. \ref{fig:rhmetal}). These first clusters enriched the ICM of their local environment with metals, from which the somewhat younger clusters of the old population were born. An increasing metallicity with time leads to successively shorter gas expulsion time-scales (Fig. \ref{fig:taumetal}) and strong low-mass star loss in a mass-segregated cluster (Fig. \ref{fig:afeh}). The higher metallicity clusters were less compact than the previously formed clusters which had a lower metallicity (Fig. \ref{fig:densemetal}) and they suffered stronger tidal-fields than their metal-poorer counterparts (Figs. \ref{fig:afeh} and \ref{fig:rhrt}).

The pre-(GC-host-)galaxy gas cloud contracted due to self-gravitation during this process. The cloud may have become clumpy and substructures emerged. Fragmentation of the cloud into massive star cluster forming regions made the potential grainy, thereby explaining the different and on average stronger tidal-field strengths among the younger clusters. Most of these clusters ($\approx90$ per cent) will have been quickly destroyed by residual-gas expulsion \citep[e.g.][and BK07]{ll03} loosing their stars to the halo field \citep*{kb02,bkp08}. Clusters forming next to other massive, dense objects are forced to expel their gas under more extreme conditions (a stronger tidal-field) than expected from their PD Galactocentric distance.

Gas expulsion from early forming clusters may trigger the formation of new clusters (via gas compression) in their immediate surrounding perhaps leading to \textit{cluster complexes} (CCs), i.e. clusters of star clusters. Such objects were recently found in the vicinity of spiral arms in external spiral galaxies \citep[e.g.][and references therein]{nb07}. It is imaginable that similar processes were at work in the GC forming gas cloud. Observed CCs contain dozens to hundreds of star clusters and have radii up to a few hundred parsecs \citep*[e.g.][for the Antennae galaxies]{ws95}. Merging processes in the central dense parts of CCs may form more massive clusters \citep*{k98,fk02,bkf09}. Clusters in such proximity may feel their mutual tidal-forces acting on each other and giving rise to small tidal-radii (as found for some DMPP clusters in Sec. \ref{sec:results}). Pal 5 may be such a surviving merged CC.

All these effects lead to depleted MFs and lower concentrations after gas expulsion in the younger of the OH clusters. \textit{In this sense the history of events that lead to the inner GC system may involve local rapid (on time-scales of hundreds of Myr) re-arrangements of the interstellar matter superposed on the overall contraction or collapse to the final population II spheroid}. The weak anticorrelation between $[Fe/H]$ and $\dgc$ (Tab.~\ref{tab:constraints}), though not very strongly significant according to our criterion, would be consistent with this scenario since enrichment continues while the cloud collapse proceeds and more-metal rich clusters would thus typically, but not exclusively, form closer to the centre. Our ideas of the formation of the Galactic GC system are summarised in Fig.~\ref{fig:galform}.

\section{Summary}
\label{sec:sum}
\citet*[MKB08]{mkb08a} showed that residual-gas expulsion models by \citet[BK07]{bk07} explain the on first sight curious trend between the global present day low-mass PDMF slope, $\alpha$, in Galactic globular clusters (GCs) and their concentration parameter, $c$, (Fig. \ref{fig:dmpp}) as found by \citet[DMPP]{dmpp07}: Weakly concentrated, low-mass-star-depleted clusters are a natural outcome of an initially mass-segregated cluster evolving through residual-gas expulsion. Given the DMPP data, it becomes possible to constrain the time-scale of gas expulsion, the tidal-field strength and the star formation efficiency (SFE) all GCs in such a sample must have had. This is achieved by direct comparison of the MKB08 models of residual-gas expulsion with the observations by DMPP and assuming the stellar low-mass IMF is universal as is indicated to be the case from a large variety of stellar systems. The above parameters determine the expansion and mass loss following the residual-gas expulsion process in $N$-body models. After applying corrections for the stellar and dynamical evolution of the clusters it becomes possible to calculate their initial half-mass radii, masses and densities directly before the residual-gas throw-out.

We find that the primordial gas in different clusters is lost on short or long time-scales, measured with respect to the initial crossing-time of the cluster. Some GCs had SFEs as low as 10 per cent, others needed 50 per cent to survive until today. Tidal-fields experienced by the clusters were weak or strong. Initial pre-cluster cloud-core masses are estimated to lie between $10^5-10^7\msun$ and cluster radii typically were smaller than $1$ pc. Average birth densities would thus have been $10^5-10^7\msun$ pc$^{-3}$.

We compiled Spearman rank-order correlation coefficients to find connections between present day and initial cluster parameters and showed that:
\begin{itemize}
 \item Present day massive clusters also had massive progenitors. Present day small-sized clusters initially also had a small half-mass radius (Tab. \ref{tab:spear}).

 \item The PDMF slope, the initial cluster size and density are functions of the PD cluster metallicity (Figs. \ref{fig:afeh}, \ref{fig:rhmetal} and \ref{fig:densemetal}).

 \item Initial half-mass radii are larger and densities are lower in higher-metallicity environments, consistent with cooling being more efficient and fragmentation into stars occurring earlier before the overall global collapse of the cluster \citep{hs09}.

 \item The MF slope trend with metallicity (Fig. \ref{fig:afeh}) could hint at a metal-dependent low-mass stellar IMF. However, the Jeans mass should be higher in low-metallicity environments and one would therefore expect a smaller fraction of low-mass stars there, contrary to observations, so that a varying low-mass part of the IMF seems not a viable explanation.

 \item We interpret the PDMF observed to be depleted in low-mass stars in higher metallicity environments (Fig. \ref{fig:afeh}) as the result of a metallicity dependent gas expulsion process: Radiation from stars couples better to high-metallicity gas so that the residual-gas from star formation is expelled more efficiently and therefore probably quicker (Fig. \ref{fig:taumetal}), similar to the metallicity dependend mass-loss rates of evolving stars \citep[e.g.][]{mok07}. This also indicates efficient, i.e. fast, gas removal in clusters with a low value of the PDMF slope. If it is true, this provides for the first time an explanation for the dependence of the PDMF on the metallicity, although it has been investigated in other works before \citep{mcc86,smcc87,dpc93}.

 \item Low-mass-star-depleted clusters have the lowest initial masses (Fig. \ref{fig:amass}) and the largest initial radii (Tab. \ref{tab:spear}). This dependency of the PDMF slope may be due to more efficient gas throw-out in low-mass clusters with a large size (eq. \ref{eq:tau}). This then leads to stronger cluster expansion following residual-gas expulsion and, in case of a primordial mass-segregated cluster, to enhanced low-mass star loss across the tidal-radius.

 \item As a result, the correlation between the density of the cluster-forming cloud-core and the PDMF slope, $\alpha$, is found to be strongest of all (Fig. \ref{fig:adense}).

 \item We find no correlation of $\alpha$ with the distance of the cluster to the Galactic Centre (Tab. \ref{tab:spear}), in contrast to the results of \citet{dpc93}.

 \item The tendency of low-mass-star-depleted clusters to be initially least dense (Fig. \ref{fig:adense}) allowed us to understand a supposed higher binary fraction in PD less-concentrated clusters: a higher fraction of unseen binaries had been arbitrarily introduced by MKB08 for weakly concentrated clusters to better match the observational data. Since the low-$\alpha$ clusters have today a low concentration (Fig. \ref{fig:dmpp}) and are initially less dense, fewer binary systems are dissolved in them \citep{k95MN} and we predict a larger fraction of binaries in PD low-$c$ clusters.
\end{itemize}

Given the many correlations with metallicity, we possibly unmasked metallicity as a main parameter in the formation process of clusters and for the residual-gas expulsion process.

Knowledge about these initial conditions of star clusters lead us to the development of a picture of the formation of the old halo GC population within the framework of the contraction of a giant pre-Milky Way (MW) gas cloud \citep*{els62}. Since the clusters of the old halo are more or less co-eval and ages are consistent with the formation from a single proto-system \citep{sw02,dea05,mf09}, the age difference must be rather small (at most a few hundred Myr). From the initial conditions we see that these clusters have formed under very different conditions: Strongly concentrated, relatively metal-poor, and thus the oldest, clusters experienced weak tidal-fields only, i.e. the overall potential was smooth in the beginning so that conditions were rather similar in the pre-MW gas cloud. Gas was blown out of the clusters on comparable time-scales, because the metallicity was low (see above). Metal enrichment by these first clusters lead to successively higher [Fe/H] in later forming clusters. During that process the cloud contracted and became grainy on about a dynamical time-scale (few hundred Myr or less), perhaps triggered by residual-gas expulsion from other clusters and gas compression.

Thus, by connecting the present-day concentration and low-mass stellar depletion of GCs it thus appears possible to uncover the evolution of the physical conditions during the first dynamical time of the MW: The environment in which clusters formed changed a lot as is seen in the stronger and also more strongly varying tidal-fields experienced by the weakly concentrated clusters that are typically more metal-rich. A larger global metallicity in clusters lead to more rapid gas blow-out and the easier loss of stars across the tidal-radius. This resulted in the younger, relatively metal-rich clusters appearing today as the low-concentration GCs with depleted low-mass stellar PDMFs.
\\\\ \textbf{Acknowledgments}\\
MM was supported for this research through a stipend from the International Max Planck Research School (IMPRS) for Astronomy and Astrophysics at the Universities of Bonn and Cologne. We thank Holger Baumgardt and Jan Pflamm-Altenburg for fruitful discussions. We thank Karl M. Menten for carefully reading the manuscript and making useful suggestions.

\bibliographystyle{mn2e}
\bibliography{biblio}
\makeatletter   \renewcommand{\@biblabel}[1]{[#1]}   \makeatother

\label{lastpage}

\end{document}